\documentclass[aps,pre,twocolumn,array,epsfig,eqsecnum]{revtex4}
\usepackage{amsmath}
\usepackage{amsfonts}

\usepackage[dvips]{graphicx}

\begin{document}   
\setlength{\parindent}{0pt}

\title{Rank dependent bounds on mixedness and entanglement for quantum teleportation}
\author{K.G Paulson, S.V.M Satyanarayana}
\address{
Department of Physics, Pondicherry University, Puducherry 605 014, India}
\date{\today}

\begin{abstract}
Entanglement and mixedness of a bipartite mixed state resource are crucial for the success of quantum teleportation. Upper bounds on measures of mixedness, namely, von Neumann entropy and linear entropy beyond which the bipartite state ceases to be useful for quantum teleportation are known in the literature. In this work, we generalize these bounds and obtain rank dependent upper bounds on von Nuemann entropy and linear entropy for an arbitrary bipartite mixed state resource. We observe that the upper bounds on measures of mixedness increase with increase in the rank. For two qubit mixed states, we obtain rank dependent lower bound on the concurrence, a measure of entanglement, below which the state is useless for quantum teleportation. Werner state, which is a fourth rank state, exhibits the theoretical upper bound on mixedness among two qubit mixed states. We construct Werner like states of lower ranks and show that these states possess the theoretical rank dependent upper bounds obtained on the measures of mixedness and theoretical rank dependent lower bounds on the concurrence.

\end{abstract}

\maketitle

\section{Introduction}
Entangled states are used as resource for quantum teleportation~\cite{Bennett1993,Popescu1994,Horodecki1996, M Nielsen}. The success of quantum teleportation is quantified by optimal teleportation fidelity. The strength of entanglement is proportional to the success of teleportation for pure entangled resources. Maximally entangled pure states give maximum optimal teleportation fidelity of unity. On the other hand, for mixed states, success of teleportation depends, in addition to strength of entanglement, on mixedness of the state~\cite{SBose2000,Paulson2014,Horodecki1999}. It was shown in~\cite{SBose2000} that, for mixed entangled states, there is an upper bound on mixedness above which the state is useless for quantum teleportation. Upper bounds on measures of mixedness, namely, von Neumann entropy and linear entropy were obtained for a general bipartite $d \times d$ state. Mixed entangled states can be classified according to their rank $r$, where the rank $r$ varies between $2$ and $d^2$ for a bipartite $d \times d$ system. In the present work, we obtain rank dependent upper bounds on measures of mixedness, above which the states of respective ranks become useless for quantum teleportation.

In our previous work~\cite{Paulson2014}, we observed that mixedness and entanglement of a mixed state resource independently influence teleportation. For a state with a fixed value of mixedness, the state being entangled is only necessary for it to be a resource for quantum teleportation, but not sufficient. States with low mixedness and high entanglement turn out to be ideal resources for quantum teleportation. We argued based on the numerical work on a class of maximally entangled mixed states that there exists a rank dependent lower bound on a measure of entanglement such as concurrence, below which the states are useless for quantum teleportation. In this article, we derive rank dependent lower bounds on concurrence for a bipartite $2 \times 2$ systems.

Werner state defined as a probabilistic mixture of maximally entangled pure state and maximally mixed separable state, exhibits highest mixedness for a given optimum teleportation fidelity among all the two qubit mixed entangled states known in the literature. Werner state~\cite{Werner1989} is a rank 4 state. In the present study, we generalize the construction and obtain second and third rank Werner states. We show that rank dependent Werner states exhibit the respective rank dependent bounds obtained on measures of mixedness and entanglement for a given value of teleportaion fidelity.

\section{Upper bounds on measures of mixedness}
We consider two measures of mixedness of a state $\rho$, namely, von Nuemann entropy defined as  $S(\rho)=-Tr(\rho\ln\rho)$~\cite{J vonNeumann} and linear entropy given as $S_{L}=\frac{d}{d-1}[1-Tr(\rho^{2})]$. Further, the maximum achievable teleportaion fidelity of a bipartite $d\times d$ system in the standard teleportation~\cite{Horodecki1999} scheme is $F=\frac {fd+1}{d+1}$, where $f$ is the singlet fraction of $\rho$ given by $f(\rho)=max_{\psi}\langle\psi\vert\rho\vert\psi\rangle$. Here the maximum is over all maximally entangled states. And maximum fidelity achieved classically is $F_{cl}=\frac{2}{d+1}$. This shows that the state is useful quantum teleportation for $f>\frac{1}{d}$ ($F(\rho) >  F_{cl}$). Let $\rho_r$ denote a bipartite mixed state of $d\times d$ system whose rank is $r$, $2 \le r \le r_{max}=d^2$.  Firstly, we prove a theorem that gives rank dependent upper bounds on von Neumann entropy, above which the state is useless for quantum teleportation.\newline

\textit{Theorem}: If the entropy $S(\rho_{r})$ of a given state  $\rho_{r}$ of rank  $r$ of $d\times d$ system exceeds $\ln\sqrt{r_{max}}+(1-\frac{1}{\sqrt{r_{max}}})\ln\frac{r-1}{\sqrt{r_{max}}-1})$, then the state is not useful for quantum teleportation.\newline
\textit{Proof}: We know for a given state $\rho_r$ of $d \times d$ systems $f(\rho_r) > \frac{1}{d}$ implies the state is useful for quantum teleportaion. We have state $\rho_{r}$ of rank $r$ of $ d \times d$ systems as,
\begin{equation}
\rho_{r}=\sum_{i=1,j=1}^{r_{max}}C_{ij}\vert i\rangle\langle j\vert
\end{equation}
$\{\vert i\rangle\}$ is the basis formed from $ r_{max}$ maximally entangled states. We have from the definition of singlet fraction the largest element, say $C_{11}$ of $C_{ii}'^{s}$ is greater than or equal to $\frac{1}{\sqrt{r_{max}}}$. And it is known that Von Neumann entropy $S(\rho_{r})$ of a given state $\rho_{r}$ is less than or equal to Shannon entropy. \textit{i.e.,}
\begin{equation}\label{shannon}
S(\rho_r)\leq\sum_{i=1}^{r_{max}} C_{ii}\ln C_{ii}
\end{equation}
Shannon entropy in eq.(\ref{shannon}) is maximum for $C_{11}=\frac{1}{\sqrt{r_{max}}}$ and rest of $r-1$ elements are equal, subjected to the constraint $C_{11} >\frac{1}{\sqrt{r_{max}}}$, we get the upper bound as
\begin{widetext}
\begin{equation}
S^{\star}(\rho_r)=\sum _{i=1}^{r_{max}}C_{ii} \ln C_{ii}=\frac{\ln\sqrt{r_{max}}}{\sqrt{r_{max}}}-\left(1-\frac{1}{\sqrt{r_{max}}}\right)\ln\left(\frac{1}{r-1}\left(1-\frac{1}{\sqrt{r_{max}}}\right)\right)
\end{equation}
\end{widetext}
This implies,
\begin{equation}\label{von}
S(\rho_{r})\leq \ln\sqrt{r_{max}}+\left(1-\frac{1}{r_{max}}\right)\ln\frac{r-1}{\sqrt{r_{max}}-1}
\end{equation}
This shows that for $S(\rho_{r})$ satisfying eq.(\ref{von}), the singlet fraction is greater than $\frac{1}{\sqrt{r_{max}}}$. Thus, we prove if $S(\rho_{r})> \ln\sqrt{r_{max}}+(1-\frac{1}{r_{max}})\ln\frac{r-1}{\sqrt{r_{max}}-1}$, state $\rho_{r}$  is useless for quantum teleportation.

We also obtain an analytical expression for the rank dependant upper bound on linear entropy as a measure of mixedness as
\begin{equation}\label{linbound}
S_{L}^{\star}(\rho_{r})=\frac{r\left(r_{max}-1\right)-2\sqrt{r_{max}}\left(\sqrt{r_{max}}-1\right)}{\left(r_{max}-1\right)\left(r-1\right)}
\end{equation}
For $d=2$, we have
\begin{equation}\label{d2slub}
S_{L}^{\star}(\rho_{r})=\frac{3r-4}{3(r-1)}
\end{equation}
where $r$ is the rank of the state, which varies from 2 to 4. In our previous work~\cite{Paulson2014}, based on the analysis of a class of maximally entangled mixed states of $2 \times 2$ bipartite system given in~\cite{Ishizaka2000}, we observed that for a given value of linear entropy, there exists a rank dependent upper bound on the optimal teleportation fidelity and the upper bound increases with increase in the rank. This is equivalent to stating that for a given value of optimal telportation fidelity, there exists a rank dependent upper bound on linear entropy and the upper bound increases with rank. The above result allows us to calculate the upper bounds explicitly. If the optimal teleportation fidelity is fixed as $2/3$, the classical fidelity, the upper bound on linear entropy for states of second, third and fourth ranks are $2/3$, $5/6$ and $8/9$ respectively. States with linear entropy above the respective rank dependent upper bounds are useless for teleportation.

A class of maximally entangled mixed states (MEMS) is constructed by Ishizaka et. al~\cite{Ishizaka2000,Frank2001} and the construction is as follows.
\begin{equation}
\rho_=\lambda_{1}\vert\psi^{-}\rangle\langle\psi^{-}\vert+\lambda_{2}\vert00\rangle\langle00\vert+\lambda_{3}\vert\psi^{+}\rangle\langle\psi^{+}\vert+\lambda_{4}\vert11\rangle\langle11\vert
\end{equation}
Where $\lambda_{i}$  are the eigenvalues of the state $\rho$ $(\lambda_{1}\geq \lambda_{2}\geq \lambda_{3}\geq \lambda_{4}$ and $\lambda_1+\lambda_2+\lambda_3+\lambda_4=1)$. The Werner state, which is a convex sum of maximally entangled pure state and maximally mixed separable state, given by
\begin{equation}\label{werner}
W_4 = (1-p)\frac{I_4}{4} + p \vert \psi^{+} \rangle \langle \psi^{+} \vert
\end{equation}
Werner state corresponds to a choice of eigenvalues given by  $\lambda_{1}=\frac{1+3p}{4}$ and $\lambda_{2}=\lambda_{3}=\lambda_{4}=\frac{1-p}{4}$. Werner state is a state of rank 4. The singlet fraction for Werner state is $\frac{1+3p}{4}$ and linear entropy is estimated as $1-p^{2}$. The value of linear entropy at which fidelity is equal to classical limit can be found as $\frac{8}{9}$, which is same as the upper bound obtained above for states of rank 4. This correspondence motivated us to construct Werner like states of ranks 2 and 3. A rank $3$ Werner state can be constructed by using eigenvalues as $\lambda_{1}=\frac{1+2p}{3}$,$\lambda_{2}=\lambda_{3}=\frac{1-p}{3}$ and $\lambda_{4}=0$. We have $f(W_{3})=\frac{1+2p}{3}$ and linear entropy is equal to $S_L(W_3)=\frac{8(1-p^{2})}{9}$. Thus it is clear that $f(W_{3})\leq \frac{1}{2}$ for a value of linear entropy greater than $\frac{5}{6}$, which is the obtained upper bound for rank 3 states. To construct  rank $2$ Werner state we take $\lambda_{1}=\frac{1+p}{2}$, $\lambda_{2}=\frac{1-p}{2}$ and $\lambda_{3}=\lambda_{4}=0$. We get singlet fraction and linear entropy as $\frac{1+p}{2}$ and $\frac{2(1-p^{2})}{3}$ respectively. It clearly shows that rank $2$ Werner state is useless for teleportation when linear entropy is greater than $\frac{2}{3}$, coinciding with the theoretical upper bound shown above. Thus, we illustrate that the constructed rank dependent Werner states exhibit the theoretical upper bounds on the linear entropy for a given value of optimal teleportaion fidelity.\newline
\section{Lower bounds on measure of entanglement}
The relationship between the concurrence as a measure of  entanglement of state $\rho$  and it's purity is well studied,  the degree of entanglement decreases as purity decreases. The maximum possible value of concurrence  of a state $\rho$ for a spectrum of eigenvalues ~\cite{Wooters2001} $(\lambda_{1},\lambda_{2},\lambda_{3},\lambda_{4})$  in  descending order is given as
\begin{equation}\label{concurrence}
C_{max}{(\rho)}=max(0,\lambda_{1}-\lambda_{3}-2\sqrt{\lambda_{2}\lambda_{4}})
\end{equation}
In our previous work, we also observed that there exist a rank dependent lower bound on  concurrence for a fixed value of optimal teleportation fidelity and the lower bound decreases with increase in the rank of the state. In the present work, we obtain the exact rank dependent lower bounds on the concurrence for a $2 \times 2$ bipartite states. We find that the eigenvalues $\lambda_{1}=\frac{1}{\sqrt{r_{max}}}$ and equal values of rest of the $(r-1)$ $\lambda_{i}'^s$  minimize  $C_{max}$ among all  eigenvalue spectra.

By substituting the values for $\lambda_{i}^{'s}$, that is, corresponding to rank $r$, $\lambda_{1}=\frac{1}{2}$ and rest of $(r-1)$ elements are equal, we get the lower bound on the concurrence of state $\rho_{r}$ of rank $r$ below which the state fails to be a source for quantum teleportation. We obtain the values of lower bound on concurrence for the failure of teleportation for rank 4,3 and 2 states as follows, $C_{4}=0, C_{3}=\frac{1}{4}$ and $C_{2}=\frac{1}{2}$ respectively.  From this we clearly show that there exist a lower bound on concurrence for a fixed value of fidelity, this lower bound decreases as rank increases for two qubit systems. The lower bound on concurrence for failure of quantum teleportation for rank dependent Werners states coincide with the bounds derived in this work.\newline

\section{Discussion}

Lower and upper bounds on teleportation fidelity as a function of concurrence are obtained in~\cite{Frank2002}. The upper and lower bounds on fidelity $F$ for a concurrence $C$ are given by $F\leq\frac{2+C}{3}$ and $F\geq max(\frac{3+C}{6},\frac{2C+1}{3})$ respectively. The upper bound on fidelity as a function of concurrence coincides with Werner state of rank 4. Based on the properties of lower rank Werner states constructed in this work, we conjecture that there exist rank dependant bounds on fidelity as a function of concurrence and these bounds which also coincide with Werner states of lower ranks. The maximum amount of fidelity of a rank three state as a function of concurrence is given as $F\leq(\frac{5+4C}{9})$, in the same way the amount of teleportation fidelity for rank two state is given by Werner state of rank 2 as $F\leq(\frac{2+C}{3})$.

\begin{figure}
  \includegraphics[width=0.5\textwidth]{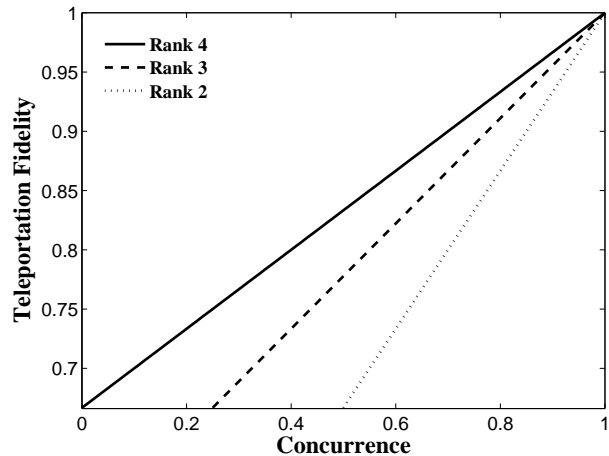}
\caption{Teleportation fidelity as a function of concurrence for rank dependent Werner states}
\label{fig:1}       
\end{figure}

Teleportation fidelity as a function of concurrence for rank dependent Werner states are presented in fig.~\ref{fig:1}. Werner states of different ranks exhibit respective rank dependent lower bounds on concurrence below which the state is useless for quantum teleportation. It can be seen that, in the fidelity - concurrence plane, the allowed values of teleportation fidelity for a fixed value of concurrence of $2 \times 2$ states is bounded below and above by curves corresponding to second and fourth rank Werner states respectively. Fidelity of second and fourth rank Werner state also coincide respectively with lower and upper bounds on fidelity of $2 \times 2$ states obtained in~\cite{Frank2002}. The curve corresponding to Werner state of rank 3, which lies between the curves of upper and lower bounds, serves as an upper bound on teleportation fidelity of rank 3 mixed states. This implies, quantum teleportation can be achieved with a resource of low value of concurrence, then it has to be a high rank state. For example, if we have to use a state with concurrence 0.1 as a quantum teleportation resource, it has to be necessarily a fourth rank state.

In~\cite{PBadziag2000} it is shown that local environment can enhance fidelity of quantum teleportation. It is shown that for a class of density matrices, interaction with environment enhance the singlet fraction above $\frac{1}{2}$, and thus making the state useful for quantum teleportation. There are other methods ~\cite{Bennet1996,Frank V2001} like entanglement purification, local filtering, entanglement concentration for single and multiple number of qubits can be made use to improve the teleportation fidelity of quantum channels. In this context it is important to understand whether the rank dependant bounds on fidelity as a function of linear entropy and concurrence are preserved under operations that enhance fidelity. We understand that the operations involved in enhancing fidelity need not preserve the rank and hence rank dependant bounds are valid as long as the state is of the respective ranks. It can be stated that for fixed values of concurrence and linear entropy of a state of fixed rank $r$ , the maximum achievable teleportation fidelity is that of Werner state of rank $r$.

We proved the existence of rank dependent upper bound on the von Neumann entropy as well as linear entropy as measures of mixedness of a general mixed state of a bipartite $d \times d$ system for failure of the state to be resource for quantum teleportation. We constructed rank 3 and rank 2 Werner states. Rank dependent Werner states exhibit the theoretical upper bounds obtained on von Neumann entropy and linear entropy. Further, we proved the existence of rank dependent lower bounds on the concurrence of a $2 \times 2$ mixed states for the failure of states as resource for quantum teleportation and showed that the lower bound on concurrence for rank dependent Werner states coincide with the theoretical lower bounds. We also showed the rank dependant Werner states give upper and lower bounds on fidelity as a function concurrence, which are consistent with the results in the literature.

\end{document}